\documentclass{aa}
\usepackage{natbib}  % added by zhijia
\usepackage{xcolor} % added by zhijia
\bibpunct{(}{)}{;}{a}{}{,} % added by zhijia
\usepackage{graphicx}
\usepackage{txfonts}
\begin{document}

   \title{Asteroseismic analysis of solar-mass subgiants KIC 6442183 and KIC 11137075 observed by \emph{Kepler}}
   \titlerunning{Asteroseismic analysis of subgiants KIC 6442183 and KIC 11137075}
   \authorrunning{Tian et al.}
   \author{Zhijia Tian
          \inst{1}
          \and
          Shaolan Bi
          \inst{1}
          \and
          Timothy R. Bedding
          \inst{2,}
          \inst{3}
          \and
          Wuming Yang
          \inst{1}
          }

   \institute{Department of Astronomy, Beijing Normal University, Beijing 100875, China; \\
              \email{tianzhijia@mail.bnu.edu.cn}; \email{bisl@bnu.edu.cn}
         \and
             Sydney Institute for Astronomy (SIfA), School of Physics, University of Sydney, NSW 2006, Australia; \\ \email{t.bedding@physics.usyd.edu.au};
         \and
             Stellar Astrophysics Centre, Department of Physics and Astronomy, Aarhus University, Ny Munkegade 120, DK-8000 Aarhus C, Denmark
             }

%   \date{Received September 15, 1996; accepted March 16, 1997}

% \abstract{}{}{}{}{}
% 5 {} token are mandatory

  \abstract
   {Asteroseismology provides a powerful way to constrain stellar parameters. Solar-like oscillations have been observed on subgiant stars with the \emph{Kepler\/} mission. The continuous and high-precision time series enables us to carry out a detailed asteroseismic study for these stars.
   }
   {We carry out data processing of two subgiants of spectral type G: KIC 6442183 and KIC 11137075 observed with the \emph{Kepler} mission, and perform seismic analysis for the two evolved stars. } % solar analog stars.
   {We estimate the values of global asteroseismic parameters: $\Delta\nu=64.9\pm 0.2 $ $\mu$Hz and $\nu_{\rm max}=1225 \pm 17$ $\mu$Hz for KIC 6442183, $\Delta\nu=65.5\pm 0.2 $ $\mu$Hz and $\nu_{\rm max}=1171 \pm 8$ $\mu$Hz for KIC 11137075, respectively. In addition, we extract the individual mode frequencies of the two stars.
   We compare stellar models and observations, including mode frequencies and mode inertias.
   The mode inertias of mixed modes, which are sensitive to the stellar interior, are used to constrain stellar models.
   We define a quantity $d\nu_{\rm m-p}$ that measures the difference between the mixed modes and the expected pure pressure modes, which is related to the inertia ratio of mixed modes to radial modes. }
   {Asteroseismic together with spectroscopic constraints provide the estimations of the stellar parameters:
   $M = 1.04_{-0.04}^{+0.01} M_{\odot}$, $R = 1.66_{-0.02}^{+0.01} R_{\odot}$ and $t=8.65_{-0.06}^{+1.12}$ Gyr for KIC 6442183, and
   $M = 1.00_{-0.01}^{+0.01} M_{\odot}$, $R = 1.63_{-0.01}^{+0.01} R_{\odot}$ and $t=10.36_{-0.20}^{+0.01}$ Gyr for KIC 11137075. Either mode inertias or $d\nu_{\rm m-p}$ could be used to constrain stellar models.}
   {}

   \keywords{Stars: oscillations --
             Stars: fundamental parameters --
             Stars: individual(KIC 6442183, KIC 11137075)
               }
   \maketitle

%
%________________________________________________________________

\section{Introduction}

\begin{table*}[!ht]%[!htb]
\begin{center}
\caption{Basic parameters of KIC 6442183 and KIC 11137075 from observations.\label{tbsp}}
\begin{tabular}{ccc}
\hline\hline
  & KIC 6442183 & KIC 11137075 \\ % & $d_{y}$ & $n$ & $\chi^2$ & $R_{maj}$ & $R_{min}$ &
\hline
$T_{\rm{eff}}$ \ (K) & $5740\pm70$\tablefootmark{a} \ \ $5738\pm62$\tablefootmark{b}   & $5590\pm70$\tablefootmark{a}  \  \   $5610\pm71$\tablefootmark{b}   \\
$\log g$ \ (dex)  & $4.03\pm0.03$\tablefootmark{a} \ \ $4.14\pm0.10$\tablefootmark{b}  & $4.01\pm0.03$\tablefootmark{a} \  \  $4.10\pm0.12$\tablefootmark{b} \\
$\rm{[Fe/H]}$   & $-0.11\pm0.06$\tablefootmark{a} \ \ $-0.12\pm0.05$\tablefootmark{b} & $-0.06\pm0.06$\tablefootmark{a} \ \ $-0.06\pm0.06$\tablefootmark{b} \\
KIC magnitude & 8.52\tablefootmark{e} & 10.86\tablefootmark{e} \\
$\Delta\nu$ \ ($\mu$Hz)   & $65.07 \pm 0.09 $\tablefootmark{c} \ \ $64.9\pm 0.2 $\tablefootmark{d}  & $65.5\pm 0.2 $\tablefootmark{d}  \\
$\nu_{\rm max}$ \ ($\mu$Hz)   & $1160  \pm 4$\tablefootmark{c} \ \ $1225  \pm 17$\tablefootmark{d}   & $1171 \pm 8$\tablefootmark{d}            \\ % &-0.4 &60 &1.4  &1.669\tablefootmark{c}
$M$ ($M_{\odot}$) & 0.94\tablefootmark{c} \ \ $1.04_{-0.04}^{+0.01}$\tablefootmark{d} & $1.00_{-0.01}^{+0.01}$ \tablefootmark{d} \\
$R$ ($R_{\odot}$) & 1.60\tablefootmark{c} \ \ $1.66_{-0.02}^{+0.01}$\tablefootmark{d} & $1.63_{-0.01}^{+0.01}$ \tablefootmark{d} \\
\hline
\end{tabular}
%% Any table notes must follow the \end{tabular} command.
\tablefoot{
\tablefoottext{a}{ \citet{Bruntt12} }
\tablefoottext{b}{ \citet{Molenda2013} }
\tablefoottext{c}{ \citet{Benomar13}, who stated that typical uncertainties on mass and radius were a few percent }
\tablefoottext{d}{This work}
\tablefoottext{e}{\emph{Kepler} Asteroseismic Science Operations Center: http://kasoc.phys.au.dk/.}
}
\end{center}
\end{table*}

  Asteroseismology provides a useful tool to probe stellar interiors, test internal physical processes and obtain stellar parameters accurately \citep[e.g.][]{Eggenberger2004AA,2008ApJ...673.1093B,Kallinger10,Montalban2013}. Many solar-like stars have been observed continuously and precisely with space missions such as \emph{CoRoT} and \emph{Kepler} \citep{Corot,Borucki2007}, ushering in a golden age for the asteroseismic analysis of oscillating stars.

The oscillations of evolved stars include mixed modes, which behave as pressure modes (p-modes) in the envelope and gravity modes (g-modes) in the core \citep{Osaki75,Aizenman77}.
Mixed modes have been used to distinguish between red clump stars and RGB stars \citep{Bedding11,Mosser11a}, and to monitor stellar evolution status from the main sequence to the asymptotic giant branch \citep{Mosser14}.
The signature of avoided crossings raises the exciting possibility that detailed modeling of the star will provide a very precise determination of its age \citep{Gilliland10,Chaplin10,Metcalfe10,Benomar12,2014aste.book...60B}.

As discussed by \citet{2001MNRAS.328..601D,2004SoPh..220..137C,2009A&A...506...57D,Benomar14,Datta15}, the g-dominated non-radial modes have much higher inertia than the radial modes, whereas the p-dominated modes do not.
Because the mixed modes possess the properties of partial g-modes, the inertia of mixed modes is usually higher than that of p-modes. The inertia of mixed modes could provide powerful constraints on stellar models.

Subgiants are the bridge between the main sequence (MS) and red giants on the H-R diagram. The study of subgiants could provide valuable insight into stellar structure and evolution.
When stars leave the main sequence stage, the g-mode and p-mode frequencies overlap, which results in the appearances of mixed modes in post-main-sequence stars.
KIC 6442183 (aka. `Dougal') and KIC 11137075 (aka. `Zebedee') are at the beginning of the subgiant stage.
The magnitudes of the stars are 8.52 and 10.86 (KIC magnitude), which are bright enough to allow seismic observations.
They have been observed by the \emph{Kepler} mission continuously with a short cadence of 58.84 seconds \citep{Gilliland10b} for quarters 6.1--17.2 and 7.1--11.3, respectively.
\citet{Benomar13,Benomar14} extracted the mode frequencies and measured the mode inertia ratio for KIC 6442183. They estimated the mass $M=0.94M_{\odot}$ and the radius $R=1.60R_{\odot}$ for KIC 6442183 using scaling relations \citep{Brown91,Kjedldsen&Bedding1995}.
With oscillation observations from \emph{Kepler} and spectroscopic observations from \citet{Bruntt12} and \citet{Molenda2013}, we can now conduct asteroseismic analyses and constrain stellar parameters for the two stars accurately.

In Section 2, we review the recent observations and extract the mode frequencies for the two stars. In Section 3, we construct stellar models and correct the near-surface term for the theoretical frequencies of the models. Seismic analyses and stellar parameters determinations are shown in Section 4. Finally, the discussions and conclusions are presented in Section 5.

\section{Observations and Data Processing}

Spectroscopic observations from \citet{Bruntt12} and \citet{Molenda2013} provided the values of metallicity ($\rm{[Fe/H]}$), effective temperature ($T_{\rm{eff}}$) and gravity ($\log g$) for these two stars. Results from the two groups show good consistency, and yield preliminary constraints on the stars. In this paper we adopt the parameters obtained by \citet{Bruntt12} because they took the asteroseismic $\log g$ into account during the spectroscopic analysis.

We obtained the power spectra by applying the Lomb-Scargle Periodogram \citep{Lomb,Scargle82} to the short cadence time series corrected by KASC Working Group 1 (WG\#1; `solar-like oscillating stars') following \citet{2011MNRAS.414L...6G}, which is available to the Kepler Asteroseismic Science Consortium \citep[KASC;][]{Kjeldsen2010} through the KASOC database \footnote{\emph{Kepler} Asteroseismic Science Operations Center: http://kasoc.phys.au.dk/.}.

\begin{figure*}[!htb]
\centering
\includegraphics[width=16cm]{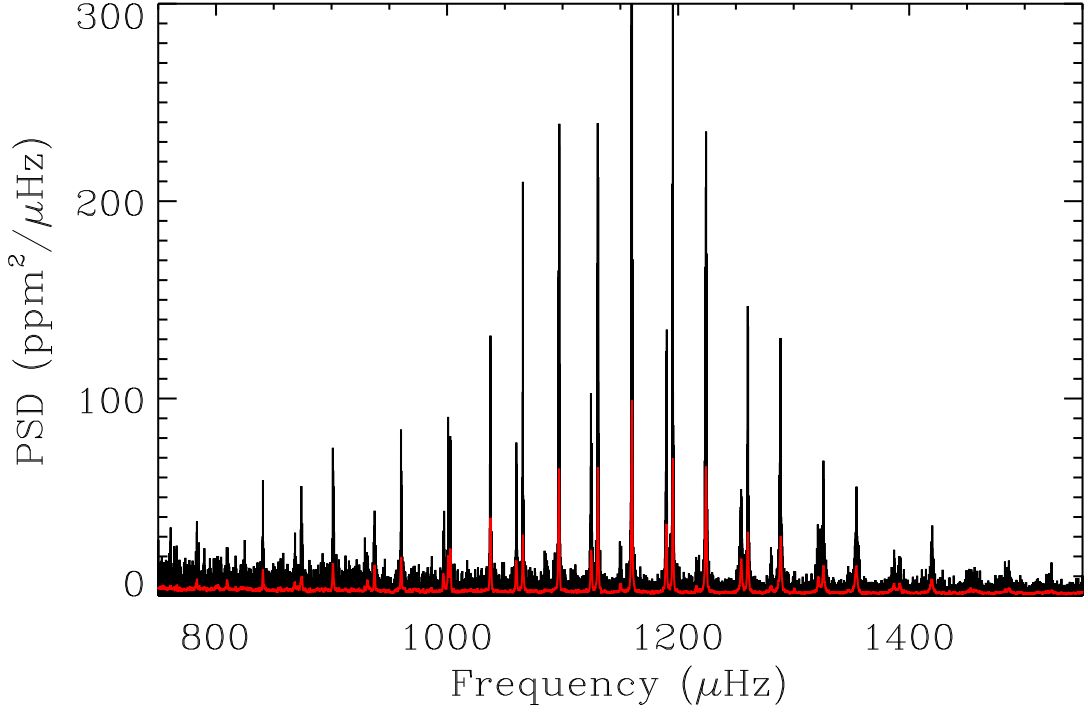}
\caption{Power spectrum for KIC 6442183. The black and red lines denote the power spectrum before and after smoothing to 2 $\mu$Hz. \label{figdd}}
\end{figure*}

\begin{figure*}[!htb]
\centering
\includegraphics[width=16cm]{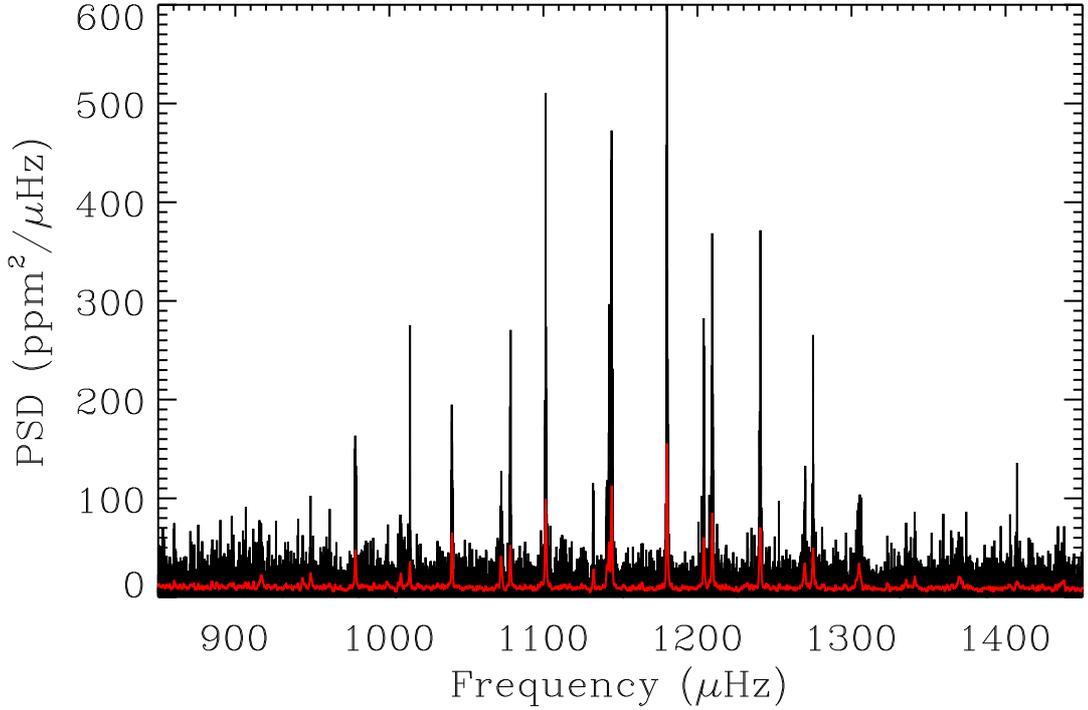}
\caption{Same as Fig. \ref{figdd} but for KIC 11137075.
\label{figzz}}
\end{figure*}

The autocorrelation of the power spectrum can provide the estimate of periodic information, such as mean large frequency separation \citep{Barban09}.
In this work, we estimated the large frequency separations $\Delta\nu$ and the frequency of maximum power $\nu_{\rm max}$ with the autocorrelation function \citep[e.g.][]{Roxburgh06,Roxburgh09,Mosser09} and collapsed autocorrelation function \citep[e.g.][]{Huber2009CoAst,Tian14}. % for the two stars.
These global oscillation parameters were evaluated as: $\Delta\nu=64.9\pm 0.2$ $\mu$Hz and $\nu_{\rm max}=1225 \pm 17$  $\mu$Hz for KIC 6442183, $\Delta\nu=65.5\pm 0.2$ $\mu$Hz and $\nu_{\rm max}= 1171 \pm 8$ $\mu$Hz for KIC 11137075.
The global oscillation parameters ($\Delta\nu$ and $\nu_{\rm max}$) and atmospheric constraints for the two stars are listed in Table$\sim$\ref{tbsp}.

As discussed by \citet{Mathur2011}, several methods have been developed to extract individual frequencies, and the key point is to fit a sum of Lorentzian profiles describing all oscillation modes.
  We followed the method described by \citet{2000ApJ...535.1078B} and \citet{2007A&A...469..233R},
  where each oscillation mode in the power spectrum was fitted to a Lorentzian profile using robust non-linear least squares method \citep{2009ASPC..411..251M,2012ascl.soft08019M}.
As discussed by \citet{1998A&AS..132..107A,1998A&AS..132..121A}, the noise statistic of each bin of the power spectrum
has a $\chi^2$ distribution with 2 degrees of freedom. However the non-linear least squares method assumes that this statistics is Gaussian.
The mode extraction method adopted in this work might introduce a certain bias of amplitudes and linewidths, and  fortunately does not affect the frequencies.
To estimate the errors, we measured the individual frequencies quarter by quarter, and estimated the frequency and error of each mode by combining each subset.
Asymptotic formulae show that the oscillation mode linewidthds are related to $T_{\rm eff}$ of stars with a power law \citep{2009A&A...500L..21C,2012A&A...537A.134A}. The smallest linewidth of the detected modes for KIC 6442183 is of approximately 0.25 $\mu$Hz \citep{Benomar13}. Because the difference of $T_{\rm eff}$ between the two subgiants is about 150 K, they will obtain the linewidths of the same magnitude. Therefore, the frequency resolution of a \emph{Kepler} quarter (0.1286 $\mu$Hz) is sufficient to resolve the oscillation modes for the two stars.

\begin{figure}[htbp]
\includegraphics[width=9.0cm]{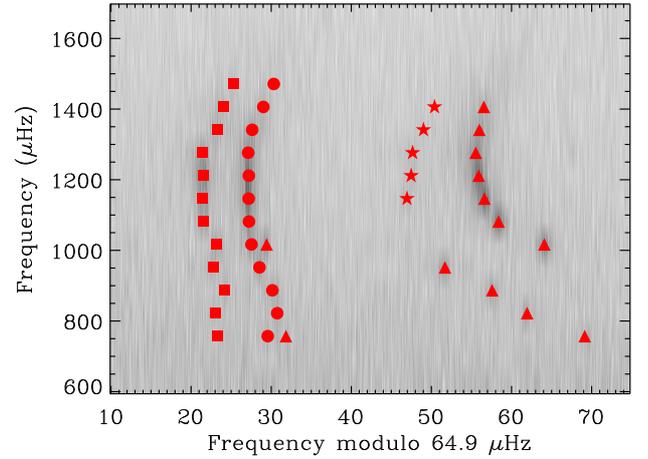}
\caption{Echelle Diagram with the identified modes for KIC 6442183. The squares denote the $l=2$ mode, while the circles $l=0$, the triangle $l=1$ and the five-pointed star $l=3$ modes. \label{figed}}
\end{figure}

\begin{figure}[htbp]
\includegraphics[width=9.0cm]{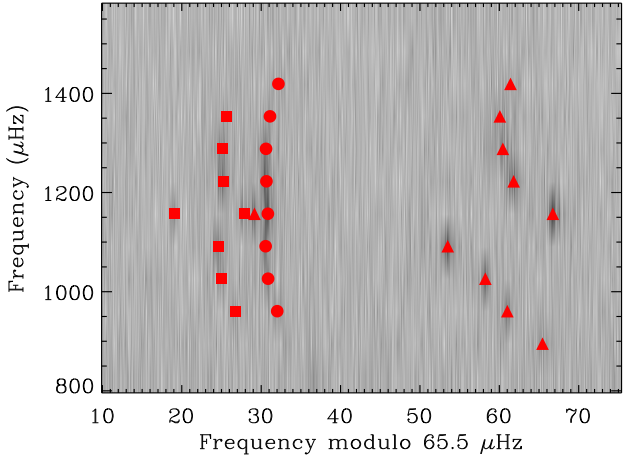}
\caption{Echelle Diagram with the identified modes for KIC 11137075. The squares denote the $l=2$ modes, while the circles $l=0$, the triangle $l=1$ modes. \label{figez}}
\end{figure}

This procedure provided us with independently obtained frequency sets with symmetric errors.
We obtained 37 oscillation modes of degree $l=0-3$ for KIC 6442183 and 26 oscillation modes of degree $l=0-2$ for KIC 11137075. The individual frequencies for the two stars are listed in Tables \ref{table:freq1} and \ref{table:freq}.
The oscillation frequencies for KIC 6442183 in this work and those given by \citet{Benomar13} are consistent at the $1-\sigma$ level.
The mode frequencies of the power spectra are over-plotted on the \'{e}chelle diagram in Figs. \ref{figed} and \ref{figez}. The radial modes follow a vertical ridge in the figures, and the non-radial modes show avoided crossings, especially for the $\emph{l} =1$ modes. In addition to the $l=0-2$ modes, there are five $l=3$ modes identified for KIC 6442183, which will provide extra constraints on theoretical models.

  \begin{table}[t]%[!htbp]
  \caption{Observed frequencies for KIC 6442183.}             % title of Table
  \label{table:freq1}      % is used to refer this table in the text
  \centering                          % used for centering table
  \begin{tabular}{l c c c}        % centered columns (4 columns)
  \hline\hline                 % inserts double horizontal lines
    \emph{l} & Frequencies ($\mu$Hz) &  error ($\mu$Hz)   \\    % table heading
  \hline                        % inserts single horizontal line
  0  &    809.63 &      0.11   \\
  0  &    874.08 &      0.25   \\
  0  &    937.21 &      0.10   \\
  0  &   1001.04 &      0.12   \\
  0  &   1065.65 &      0.14   \\
  0  &   1130.50 &      0.09   \\
  0  &   1195.40 &      0.03   \\
  0  &   1260.32 &      0.12   \\
  0  &   1325.73 &      0.19   \\
  0  &   1391.69 &      0.55   \\
\hline
  1  &    783.35 &      0.31   \\
  1  &    840.62 &      0.11   \\
  1  &    901.27 &      0.14   \\
  1  &    960.26 &      0.09   \\
  1  &   1002.91 &      0.21   \\
  1  &   1037.60 &      0.09   \\
  1  &   1096.83 &      0.25   \\
  1  &   1159.96 &      0.06   \\
  1  &   1224.09 &      0.13   \\
  1  &   1288.75 &      0.22   \\
  1  &   1354.10 &      0.26   \\
  1  &   1419.52 &      0.37   \\
\hline
  2  &   801.63 &       0.35   \\
  2  &   868.11 &       0.17   \\
  2  &   931.16 &       0.30   \\
  2  &   996.92 &       0.15   \\
  2  &   1059.95 &      0.17   \\
  2  &   1124.73 &      0.23   \\
  2  &   1189.88 &      0.15   \\
  2  &   1254.61 &      0.14   \\
  2  &   1321.26 &      0.31   \\
  2  &   1386.81 &      0.41   \\
\hline
  3  &   1150.13 &      0.25   \\
  3  &   1215.93 &      0.31   \\
  3  &   1280.49 &      0.31   \\
  3  &   1347.26 &      0.37   \\
  3  &   1412.36 &      0.59   \\

  \hline                                   %inserts single line
  \end{tabular}
  \end{table}

  \begin{table}
  \caption{Observed frequencies for KIC 11137075.}             % title of Table
  \label{table:freq}      % is used to refer this table in the text
  \centering                          % used for centering table
  \begin{tabular}{l c c c}        % centered columns (4 columns)
  \hline\hline                 % inserts double horizontal lines
    \emph{l} & Frequencies ($\mu$Hz) &  error ($\mu$Hz)   \\    % table heading
  \hline                        % inserts single horizontal line
     0  &  949.02 &      0.22      \\
     0  & 1013.37 &      0.22      \\
     0  & 1078.56 &      0.09      \\
     0  & 1144.34 &      0.10      \\
     0  & 1209.66 &      0.13      \\
     0  & 1275.11 &      0.08      \\
     0  & 1341.11 &      0.25      \\
     0  & 1407.67 &      0.45      \\
\hline
     1  &  916.94 &      0.33      \\
     1  &  978.01 &      0.11      \\
     1  & 1040.74 &      0.10      \\
     1  & 1101.50 &      0.10      \\
     1  & 1142.65 &      0.12      \\
     1  & 1180.26 &      0.09      \\
     1  & 1240.80 &      0.07      \\
     1  & 1304.94 &      0.19      \\
     1  & 1370.07 &      0.62      \\
     1  & 1436.91 &      1.05      \\
\hline
     2  &  943.76 &      0.47      \\
     2  & 1007.47 &      0.21      \\
     2  & 1072.59 &      0.21      \\
     2  & 1132.52 &      0.17      \\
     2  & 1141.38 &      0.23      \\
     2  & 1204.19 &      0.07      \\
     2  & 1269.64 &      0.20      \\
     2  & 1335.61 &      0.73      \\
  \hline                                   %inserts single line
  \end{tabular}
  \end{table}

\section{Stellar Models}
To estimate the parameters of the two stars, a grid of stellar evolutionary models was constructed with the Yale stellar evolution code \citep[YREC7;][]{2008Ap&SS.316...31D}. We used the OPAL opacity table GN93 \citep{GN93}, the low-temperature table AGS05 \citep{AGS05}, the OPAL equation-of-state tables EOS2005 \citep{eos2005} and the Bahcall nuclear rates \citep{Bahcall95} for microphysics. We chose the Eddington grey atmosphere $T-\tau$ relation. We adopted the standard mixing-length theory \citep{Bohm58} and overshooting to treat convection. The coefficient of helium and heavy elements diffusion is from \citet{Thoul94}. We did not take rotation or magnetic field into consideration in our calculations.

The ratio of heavy-elements to hydrogen as a mass fraction was estimated through the formula:
    \begin{equation}
      \rm{[Fe/H]}=\log\left(\frac{Z}{X}\right)-\log\left(\frac{Z}{X}\right)_{\odot} ,
    \end{equation}
where we adopt the value of $\left(\frac{Z}{X}\right)_{\odot} = 0.0245$ \citep{GN93} and the values of \rm{[Fe/H]} from \citet{Bruntt12} for the two stars.
The values of the ratio of the heavy element abundance to hydrogen abundance $\left(\frac{Z}{X}\right)_{s}$ are in the range of 0.0165--0.0218 and 0.0186--0.0245 for KIC 6442183 and KIC 11137075, respectively. In the model calculation, we chose the initial helium abundance as $Y_{i}=0.245+ 1.54Z_{i}$ \citep[e.g.,][]{Dotter2008,Thompson14}, in terms of the initial metal abundance.
For a given mass, stellar evolutionary models depend on three free parameters: initial chemical compositions, the mixing-length parameter $\alpha$ and the overshooting parameter $\alpha_{ov}$.
The input parameters for model calculations are listed in Table$\sim$\ref{tbinput}.

\begin{table}
\begin{center}
\caption{Input parameters for model calculations.\label{tbinput}}
\begin{tabular}{lccccrrrr}
\hline\hline
  Variable & Range &  $\delta$   \\
\hline
 Mass $(M_{\odot})$   & 0.90 -- 1.08    &  0.01       \\
     $Z_{i}$          & 0.012 -- 0.018  &  0.001      \\
     $\alpha$         & 1.70 -- 1.90    &  0.2        \\
     $\alpha_{ov}$    & 0.0 -- 0.2    &  0.2        \\
\hline
\end{tabular}
\end{center}
\end{table}

\begin{figure}
\includegraphics[width=9cm]{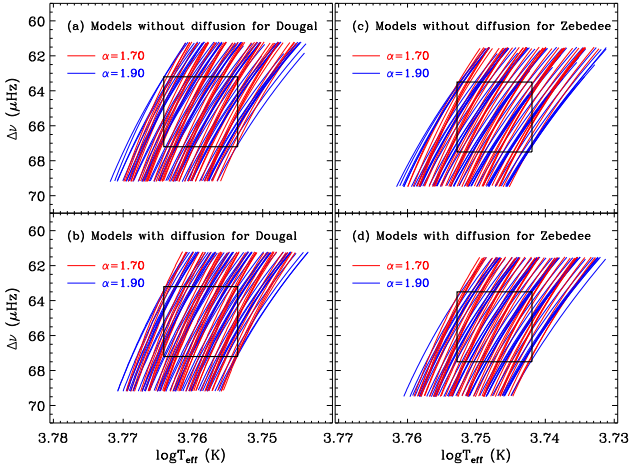}
\caption{Tracks for KIC 6442183 falling into the error box in the H-R diagram. The left and right panels denote models for KIC 6442183 and KIC 11137075, respectively. Red and blue colors in each panel denote models with mixing-length parameter $\alpha=1.70$ and $\alpha=1.90$, respectively. \label{hr1}}
\end{figure}

We calculated three groups of tracks for these two stars: (i) models without diffusion and without overshooting, (ii) models with diffusion and without overshooting and (iii) models with diffusion and overshooting ($\alpha_{ov}=0.2$).
Because the observed luminosities of the two stars were not available, we used the large frequency separation $\Delta\nu$ to constrain stellar models in the H-R diagram. Considering the difference between $\Delta\nu$ of the models from the scaling relation and pulsation code, we assigned the large separation an error of approximately 2 $\mu$Hz in the H-R diagram. There are 256 and 265 tracks falling into their error boxes for KIC 6442183 and KIC 11137075, respectively, as shown in Fig. \ref{hr1}.

\section{Asteroseismic Diagnostics}
\subsection{Model calibration}
For stellar models along the tracks falling inside the error box,
we computed smaller and more finely sampled grids around these models, and calculated theoretical mode frequencies with Guenther's pulsation code \citep{Guenther94}.
It is well known that there is a systematic offset between observed and computed frequencies that arises from improper modeling of the near-surface layers for both the Sun and solar-type stars
\citep[][and references therein]{Kjeldsen08}.
To correct the near-surface term, we followed the method described by \citet{Brandao11}, who corrected the near-surface term for radial modes with the method proposed by \citet{Kjeldsen08} by fitting a power law:
    \begin{equation}
      \nu_{\rm obs}(n,0)-\nu_{\rm best}(n,0)=a\left(\frac{\nu_{\rm obs}(n,0)}{\nu_{0}}\right)^b .
    \end{equation}
For the mixed modes, the corrected frequencies $\nu_{\rm corr}(n,l)$ could be calculated through:
    \begin{equation}
      \nu_{\rm{corr}}(n,l)=\nu_{\rm best}(n,l)+a\left(\frac{1}{Q_{nl}}\right)\left(\frac{\nu_{\rm obs}(n,l)}{\nu_{0}}\right)^b,
    \end{equation}
where $n$ and $l$ are the order and degree of the modes, respectively, $b$ is the exponent calibrated from the Solar models, and  $a$ is a constant for the stellar model. The quantity $Q_{nl}$ presents the ratio between the inertia of the non-radial mode $I_{l}$ and the inertia of a radial mode %$\varepsilon_{0}$
 $I_{0}$ of the same frequency \citep{Aerts10}. The inertia of mode $I_{l}$ is defined as:
    \begin{equation}
      \label{Il}
      I_{l}=4\pi\int_{0}^{R}[|\xi_{r}(r)|^{2}+l(l+1)|\xi_{h}(r)|^{2}]\rho_{0}r^{2}dr ,
    \end{equation}
where $\xi_{r}(r)$ and $\xi_{h}(r)$ are the radial and horizontal displacements of the modes at the radius $r$, respectively.
The inertia ratio between non-radial and radial modes can be expressed as
 \begin{equation}
  Q_{nl}=I_{l}/I_{0} \simeq 1+\frac{l(l+1)\int_{0}^{R}|\xi_{h}(r)|^{2}\rho_{0}r^{2}dr}{\int_{0}^{R}|\xi_{r}(r)|^{2}\rho_{0}r^{2}dr}.
 \end{equation}
As discussed by \citet{Aerts10}, the typical values for $\xi_{h}(R)/\xi_{r}(R)$ are 0.01--0.1 for low-order p-modes, and 10--100 for high-order g-modes. This indicates that $Q_{nl}\simeq 1$ for pure acoustic modes and $Q_{nl} > 1$ for p-g mixed modes.

To restrict the stellar parameters, we performed a $\chi_C^2$ minimization by a comparison of models with observations:
    \begin{equation}
      \chi_C^2 = \frac{1}{3}\sum_{i=1}^3 \left(\frac{C_{i}^{\rm theo}-C_{i}^{\rm obs}}{\sigma_{C_{i}}^{\rm obs}}\right)^2 ,
    \end{equation}
where $\textit{C}=(T_{\rm{eff}},\rm{[Fe/H]},\Delta\nu)$ and the $\sigma_{C_{i}}^{\rm obs}$ denote the observational errors. We chose models with $\chi_C^2<1$ as candidates for further pulsation analysis.
In addition to the constraints from the atmospheric parameters and $\Delta\nu$, we performed another $\chi_\nu^2$ minimization by a comparison of the near-surface-corrected model frequencies with the observed frequencies:
    \begin{equation}
      \chi_\nu^2=\frac{1}{N}\sum_{i=1}^N \left(\frac{\nu_{i}^{\rm theo}-\nu_{i}^{\rm obs}}{\sigma_{\nu_{i}}^{\rm obs}}\right)^2 ,
    \end{equation}
where the superscripts `obs' and `theo' correspond to individual frequencies from observations and modeling, respectively, and $\sigma_{\nu_{i}}^{\rm obs}$ denotes the observational errors.
We list the candidate models with $\chi_{\nu,all}^{2} < 3000 $ for KIC 6442183 in Table \ref{tabled1}. Since the $l=3$ modes of KIC 11137075 were not available, we chose the candidate models with $\chi_{\nu,all}^{2} < 2000$ for this star, as listed in Table \ref{tabled1}.
Note that the Brunt-V\"{a}is\"{a}l\"{a} frequency affects the distributions of oscillation modes undergoing avoided crossings.
Models with the similar global properties can have different mixed mode frequencies due to the diverse positions of avoided crossings. This results in the large $\chi_{\nu,all}^{2}$ for some models listed in Table \ref{tabled1}.

We note that the mass of these stars is not high enough to produce a convective core in the stellar interior, and the core overshooting will not have any impact on their structure and evolution. Therefore, we do not discuss the effects of overshooting in the following part for these low mass subgiants.

\subsection{Optimal models}

\begin{table*}%[width=48cm]
\caption{The best candidate models for the two stars.
   The input parameters were: initial mass, mixing-length parameter ($\alpha$), inclusion of diffusion (${\rm Diff}$) and initial metal abundance ($Z_{i}$).
   The model parameters were: age ($t$), surface abundance ratio ($(Z/X)_{s}$), effective temperature ($T_{\rm{eff}}$), luminosity ($L$), radius ($R$), gravity ($\log g$), mean large separation($\Delta\nu$), $\chi_C^2$, $\chi_{\nu,l=0}^2$  and $\chi_{\nu,all}^2$ (columns 6-15).}
\label{tabled1}    % is used to refer this table in the text
\centering                          % used for centering table
\begin{tabular}{l c c c c c c c c c c c c r r}        % centered columns (4 columns)
\hline
$\emph{Model}$ & $\emph{Mass}$ & $\alpha$ & ${\rm Diff}$ &   $Z_{i}$ & $\emph{t}$ & $(Z/X)_{s}$   & $T_{\rm{eff}}$ & $L$ & $R$ & $\log g$ & $\Delta\nu$ & $\chi_C^2$ & $\chi_{\nu,l=0}^2$  & $\chi_{\nu,all}^2$\\
  & ($M_{\odot}$) & & &  & (Gyr) &  & (K) & $ (L_{\odot}$) & ($R_{\odot}$) &  (dex) & ($\mu$Hz) & \\
\hline
& &  \multicolumn{10}{c}{KIC 6442183} & & \\ % KIC 6442183 \\
\hline
  1 & 0.98 & 1.70 & No & 0.012 & 10.06 & 0.017 & 5703 & 2.51 & 1.63 & 4.01 & 64.86 & 0.44 & 29.85 & 2514.27 \\
  2 & 0.99 & 1.70 & No & 0.013 & 9.95  & 0.018 & 5685 & 2.50 & 1.63 & 4.01 & 64.85 & 0.27 & 28.32 & 599.80 \\
 \textbf{3}& 1.00 & 1.70 & No & 0.014 & 9.80 & 0.019 & 5672 & 2.49 & 1.64 & 4.01 & 64.85 & 0.34 & 27.24 & 77.20 \\
 \textbf{4}& 1.04 & 1.90 & No & 0.015 & 8.62 & 0.021 & 5807 & 2.82 & 1.66 & 4.01 & 64.85 & 0.49 & 26.46 & 70.98 \\
\hline
 5 & 1.03 & 1.90 & Yes & 0.016 & 8.85 & 0.019 & 5709 & 2.62 & 1.66 & 4.01 & 64.84 & 0.09 & 22.40 & 1429.14 \\
 6 & 1.04 & 1.90 & Yes & 0.015 & 8.30 & 0.017 & 5783 & 2.77 & 1.66 & 4.01 & 64.84 & 0.31 & 27.61 & 2208.02 \\
 7 & 1.04 & 1.90 & Yes & 0.016 & 8.50 & 0.019 & 5743 & 2.69 & 1.66 & 4.01 & 64.84 & 0.03 & 23.46 & 532.24 \\
\textbf{8}& 1.04 & 1.90 & Yes & 0.017 & 8.68 & 0.020 & 5707 & 2.63 & 1.66 & 4.01 & 64.84 & 0.15 & 22.04 & 46.23 \\
 9 & 1.04 & 1.90 & Yes & 0.018 & 8.85 & 0.021 & 5673 & 2.57 & 1.66 & 4.01 & 64.84 & 0.57 & 20.54 & 571.86 \\
10 & 1.05 & 1.90 & Yes & 0.018 & 8.50 & 0.021 & 5707 & 2.65 & 1.67 & 4.02 & 64.85 & 0.32 & 23.59 & 1480.05 \\
\hline
\hline
 & &  \multicolumn{10}{c}{ KIC 11137075} & & \\
\hline
  1 & 0.96 & 1.70 & No & 0.014 & 11.60 & 0.019 & 5549 & 2.19 & 1.60 & 4.01 & 65.50 & 0.26 & 81.36 & 1292.34 \\
  2 & 1.00 & 1.90 & No & 0.015 & 10.14 & 0.021 & 5686 & 2.49 & 1.63 & 4.02 & 65.51 & 0.64 & 64.19 & 470.62 \\
 \textbf{3}& 1.00 & 1.90 & No & 0.016 & 10.39 & 0.022 & 5645 & 2.42 & 1.63 & 4.01 & 65.52 & 0.26 & 43.86 &  39.74 \\
  4 & 1.01 & 1.90 & No & 0.017 & 10.17 & 0.024 & 5645 & 2.43 & 1.63 & 4.02 & 65.53 & 0.44 & 19.19 & 839.92 \\
\hline
 \textbf{5}& 0.99 & 1.90 & Yes & 0.016 & 10.37 & 0.019 & 5587 & 2.30 & 1.62 & 4.01 & 65.53 & 0.25 & 19.16 & 111.69 \\
  6 & 1.00 & 1.90 & Yes & 0.016 & 9.96  & 0.019 & 5622 & 2.38 & 1.63 & 4.01 & 65.48 & 0.32 & 135.53 & 1712.46 \\
  7 & 1.00 & 1.90 & Yes & 0.017 & 10.16 & 0.020 & 5586 & 2.32 & 1.63 & 4.01 & 65.50 & 0.05 & 65.69 & 431.52 \\
 \textbf{8}& 1.00 & 1.90 & Yes & 0.018 & 10.36 & 0.022 & 5553 & 2.26 & 1.63 & 4.01 & 65.51 & 0.10 & 44.50 &  32.11 \\
\hline\hline
\end{tabular}
\end{table*}

The frequency separation between consecutive modes varies rapidly during an avoided crossing and so the mixed modes together with pure p-modes provide an important constraint to the stellar evolutionary state \citep{1995ApJ...443L..29C,2004MNRAS.350.1022P}.
We compared the frequencies and inertia ratios from observations and theoretical models listed in Table \ref{tabled1}.
The theoretical inertias for models were calculated using equation \ref{Il}, and the observed inertia ratio through the following function \citep[e.g.][]{Benomar14}:
\begin{equation}
  \label{I10}
   \frac{I_{1}}{I_{0}}= V_{1}\frac{A_{0}}{A_{1}}\sqrt{\frac{\Gamma_{0}}{\Gamma_{1}}},
\end{equation}
where $A$ and $\Gamma$ denote the amplitude and linewidth of modes, respectively, while the visibility $V_{1}$ is the square root of the ratio between theoretical heights for the dipole and radial modes. We adopted the observed amplitude, linewidth and the value of $V_{1}$ ($V_{1}^2$=1.52) from \citet{Benomar13} to deduce the observed inertia ratios.

In solar-like stars, p-mode oscillations are expected to follow the approximate relation \citep{Tassoul80}:
\begin{equation}
  \label{dnu}
  \nu_{n,l} \approx (n+\frac{l}{2}+\epsilon)\Delta\nu-l(l+1)D_{0} ,
\end{equation}
where $D_{0}$ is related to the interior structure of the star and the offset $\epsilon$ is sensitive to the surface layers.
The frequencies of the p-modes with the same degree show vertical  ridges in the \'{e}chelle diagram.
The mean distance between dipole modes and radial modes with the same order would be $0.5\Delta\nu-2D_{0}$,
which is deduced from the approximate relation formulated in equation \ref{dnu}.
For red giants, \citet{Mosser11} introduced the expression for the pure p-mode eigenfrequency pattern:
\begin{equation}
  \label{dnm}
  \nu_{n_{\rm p},l} = (n_{\rm p}+\frac{l}{2}+\epsilon)\Delta\nu-l(l+1)D_{0} + \frac{\alpha}{2}(n_{\rm p}-n_{\rm max})^{2}\Delta\nu ,
\end{equation}
where $n_{\rm p}$ is the p-mode radial order, $n_{\rm max}$ is the order of the mode with highest power and $\alpha$ is a constant representing the mean curvature of the p-mode oscillation pattern.
Although the dipole modes are p-g mixed modes in subgiants, we could estimate the expected pure acoustic modes  $\nu_{n_{\rm p},1} = \nu_{n_{\rm p},0}+0.5\Delta\nu-2D_{0}$ through equations \ref{dnu} or \ref{dnm}.

We define a quantity $d\nu_{m-p}$ to measure the frequency difference between the mixed mode frequency $\nu_{n,1}$ and the pure acoustic modes $\nu_{n_{\rm p},1}$.
The quantity $d\nu_{\rm m-p}$ for dipole modes could be calculated using the following formula:
\begin{equation}
  \label{dmp}
     d\nu_{\rm m-p} = |\nu_{n,1}- (\nu_{n_{\rm p},0}+0.5\Delta\nu-2D_{0}) | .
\end{equation}
The $d\nu_{\rm m-p}$ is the frequency difference between the mixed modes and the nearest p-mode.
For g-dominated mixed modes, the bigger is the $Q_{nl}$, the stronger is the coupling between p- and g-modes.
We used inertia ratios and $d\nu_{\rm m-p}$ as criteria to constrain the stellar models of KIC 6442183 and KIC 11137075.

Through the comparisons of theoretical and observed inertia ratios and $d\nu_{m-p}$ for KIC 6442183, we found that Models 3, 4 without diffusion and Model 8 with diffusion match the observations better than the other models.
We estimated the following parameters for KIC 6442183: $M = 1.04_{-0.04}^{+0.01} M_{\odot}$, $R = 1.66_{-0.02}^{+0.01} R_{\odot}$ and $t=8.65_{-0.06}^{+1.12}$ Gyr. By comparing Models 4 and 6, we found that the $T_{\rm{eff}}$ and luminosity estimated from the model without diffusion are higher than that estimated from diffusion models, which suggests that helium and heavy metal diffusions are not negligible.
This is because the diffusion changes the opacity and equation of state for the models.
The mass we find is approximately 11\% higher than that deduced by \citet{Benomar14} with the scaling relation.

\begin{figure}
\includegraphics[width=9cm]{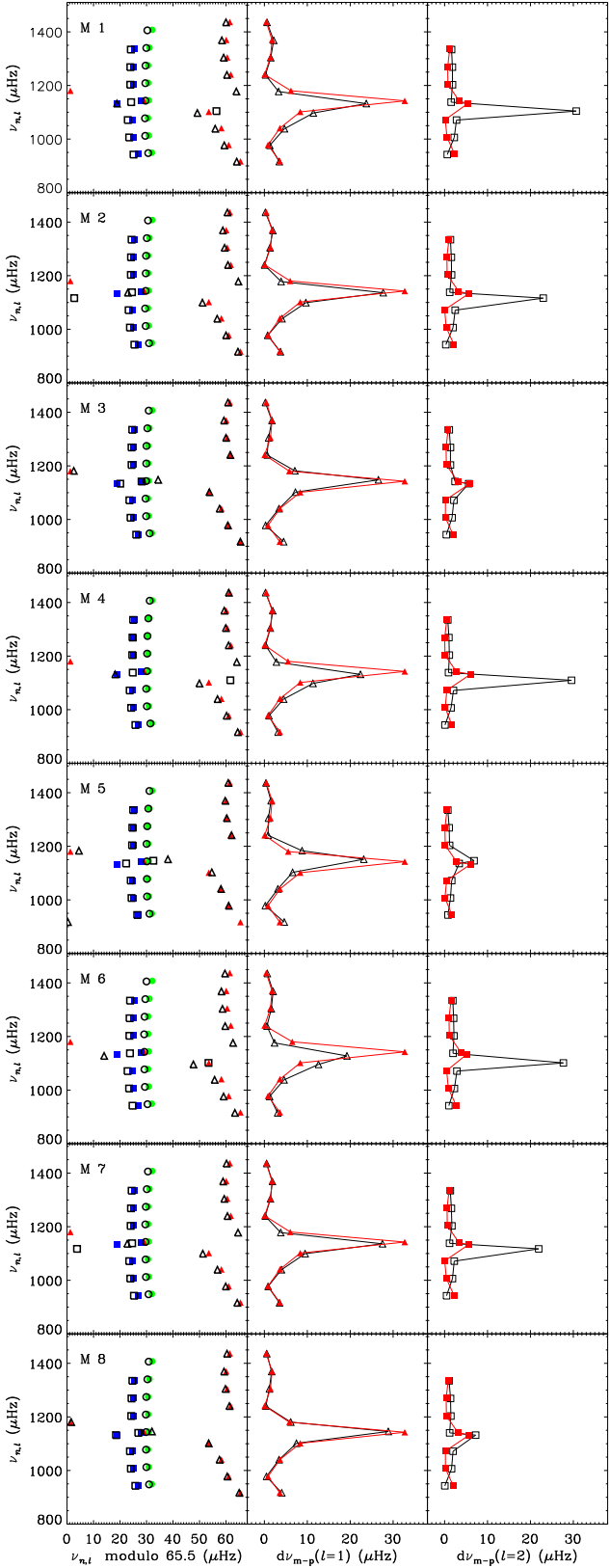}
\caption{Comparison of echelle diagram and $d\nu_{m-p}$ for observations and modeling for KIC 11137075.
The color filled symbols denote the observed frequencies, and the unfilled symbols are the corrected theoretical frequencies.
The squares denote the $l=2$ modes, while the circles show $l=0$, the triangles show $l=1$ modes. \label{ZE1}}
\end{figure}

In Fig. \ref{ZE1}, we show the results for KIC 11137075. As shown in the figure, the maximum of theoretical dipole $d\nu_{\rm m-p}$ is lower than the observations. For the $l=2$ models presented in the third column of Fig. \ref{ZE1}, Models 3, 5 without diffusion and Model 8 with diffusion match the observations much better than the other four models, which means that the three models reproduce the $l=2$ avoided crossing better than the other four models. Thus, Models 3, 5 and 8 were selected as the best models for KIC 11137075.
Finally, we obtained the following parameters for KIC 11137075:
$M = 1.00_{-0.01}^{+0.01} M_{\odot}$, $R = 1.63_{-0.01}^{+0.01} R_{\odot}$ and $t=10.36_{-0.20}^{+0.01}$ Gyr.
The models with the mixing length $\alpha =1.90$ reproduce the observations better than the models with $\alpha =1.70$.

\begin{figure}
\includegraphics[width=9cm]{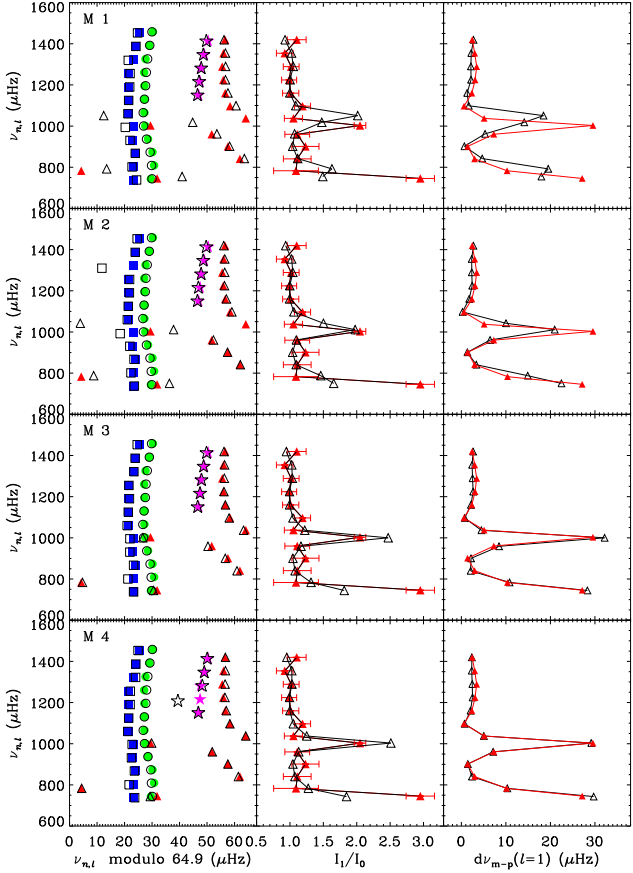}
\caption{Comparison of echelle diagram, inertia ratios and $d\nu_{m-p}$ from observations and models without diffusion for KIC 6442183.
The color filled symbols denote the observed frequencies, and the unfilled symbols are the corrected theoretical frequencies.
The squares denote the $l=2$ modes, while the circles show $l=0$, the triangle show $l=1$ and the five-pointed star $l=3$ modes.
We did not plot the error bars of frequencies since these are smaller than the symbol signs. \label{DE1}}
\end{figure}

\begin{figure}
\includegraphics[width=9cm]{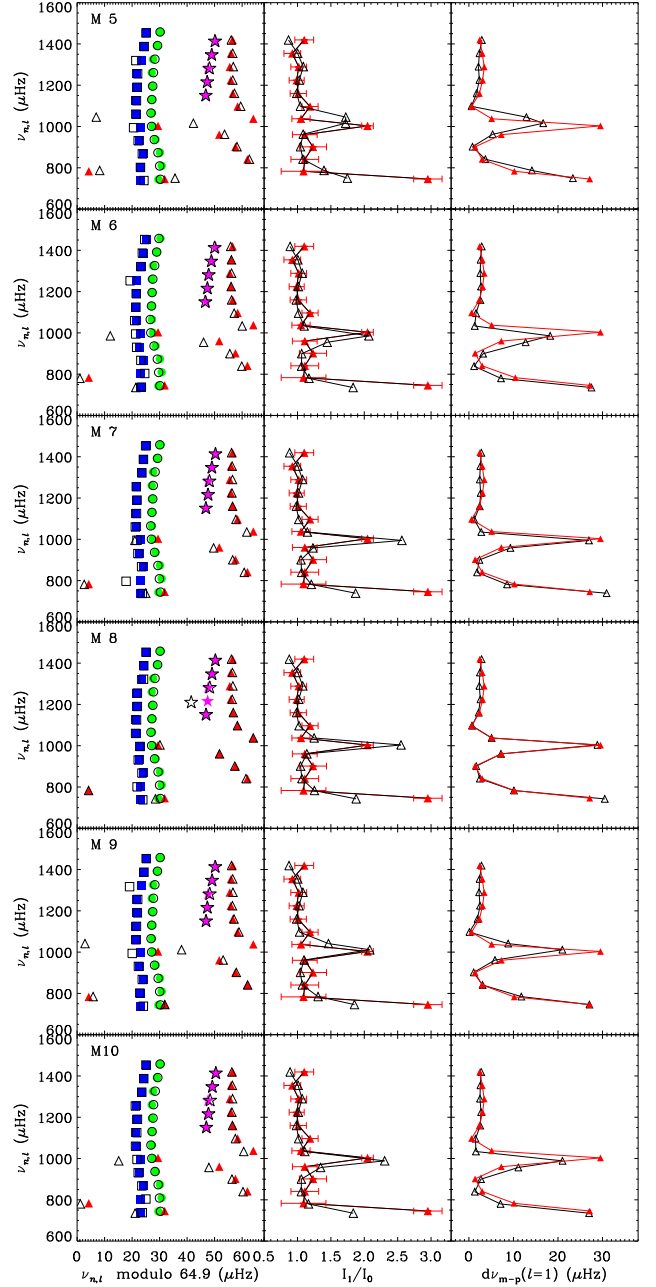}
\caption{Continuation of Fig. \ref{DE1}. \label{DE2}}
\end{figure}

\section{Discussions and Conclusions}
We carried out data processing and performed seismic analysis for subgiants KIC 6442183 and KIC 11137075, which were observed by the \emph{Kepler} Mission. The main results and discussions are summarized as follows.

    We applied the Lomb-Scargle Periodogram to the corrected short cadence time series observed by \emph{Kepler} to obtain the power spectra, and estimated the large frequency separation and frequency of maximum power for the two stars: $64.9\pm 0.2$ $\mu$Hz and $1225 \pm 17$  $\mu$Hz for KIC 6442183, $65.5\pm 0.2$ $\mu$Hz and $1171 \pm 8$ $\mu$Hz for KIC 11137075, respectively. Individual mode frequencies were also extracted in this work.
    After carrying out asteroseismic analysis, we estimated the stellar parameters: $M = 1.04_{-0.04}^{+0.01} M_{\odot}$, $R = 1.66_{-0.02}^{+0.01} R_{\odot}$ and $t=8.65_{-0.06}^{+1.12}$ Gyr for KIC 6442183, $M = 1.00_{-0.01}^{+0.01} M_{\odot}$, $R = 1.63_{-0.01}^{+0.01} R_{\odot}$ and $t=10.36_{-0.20}^{+0.01}$ Gyr for KIC 11137075.
    Both the subgiants are shown to be solar-mass stars.

    For stars whose mass is similar to the Sun's, the helium and heavy elements diffusion will affect the opacity and equation of state, which will change the thermal structure and affect the evolution of stars.
    Therefore, we have to consider the diffusion process in stellar models and seismic analysis.

    Pure p modes frequencies decrease and g modes frequencies increase with age for subgiants \citep{1995ApJ...443L..29C}, which results in the change of the frequency at which the avoided crossing occurs \citep{2014aste.book...60B}. The frequencies of mixed modes sensitive to stellar interiors are useful to constrain stellar structures, especially for stellar interiors.
    The maximum of $d\nu_{\rm m-p}$ is coincident with the large inertia ratio, allowing us to locate the mixed modes with the most g-mode property through these quantities.
    We first constrain stellar models with inertia ratios and $d\nu_{\rm m-p}$, which provides an impactful way to determine the parameters of evolved stars accurately.

    In future work, we plan to select some evolved stars with mixed modes such as subgiants, red giants and red clumps which constitute an evolution sequence. Then we compare the differences of inertia ratios and $d\nu_{\rm m-p}$ between these stars and use these quantities to constrain the stellar interiors with the quantitis.

%__________________________________________________________________

\begin{acknowledgements}
      We are grateful to the entire \emph{Kepler} team.
This work was supported by grants 10933002, 11273007 and 11273012 from the National Natural
Science Foundation of China, and the Fundamental Research Funds for the Central Universities
\end{acknowledgements}

%\bibliographystyle{aa}
%\bibliography{myref}

\end{document}